\documentstyle[12pt]{article}

\catcode`\@=11 
\def\lsim{\mathrel{\mathpalette\@versim<}}
\def\gsim{\mathrel{\mathpalette\@versim>}}
\def\@versim#1#2{\vcenter{\offinterlineskip
        \ialign{$\m@th#1\hfil##\hfil$\crcr#2\crcr\sim\crcr } }}
\catcode`\@=12 

\begin{document}
\begin{titlepage}
  \begin{flushright}
    KUNS-1473 \\[-1mm]
    HE(TH)97/16 \\[-1mm]
    hep-ph/9710417
  \end{flushright}
  \begin{center}
    \vspace*{1.2cm}
    
    {\Large\bf Neutrino Mass Texture with Large Mixing}
    \vspace{1.5cm}
    
    {\large
      Masako {\sc Bando}\footnote{E-mail address:
        bando@aichi-u.ac.jp},
      Taichiro {\sc Kugo}\footnote{E-mail address:
        kugo@gauge.scphys.kyoto-u.ac.jp}, and
      Koichi {\sc Yoshioka}\footnote{E-mail address:
        yoshioka@gauge.scphys.kyoto-u.ac.jp}}
    \vspace{7mm}

    $^*$ {\it Aichi University, Aichi 470-02, Japan} \\[1mm]
    $^{\dagger,\ddagger}$ {\it Department of Physics, Kyoto University
      Kyoto 606-01, Japan}
    \vspace{1.5cm}

    \begin{abstract}
      We propose a simple texture for the right-handed Majorana mass
      matrix to give a large $\nu_\mu\hbox{--}\nu_\tau$ mixing angle
      and hierarchical left-handed neutrino mass pattern. Consistently
      with the Dirac mass texture of the quark sector realizing the
      CKM mixing, this naturally explains the recent experimental
      results on both the atmospheric neutrino anomaly observed by the
      Superkamiokande collaboration and the solar neutrino problem. In
      this texture the right-handed Majorana mass of the third
      generation is of the order of GUT scale, which is favorable for
      reproducing the observed bottom-tau mass ratio.
      \\[3mm]
    \end{abstract}
  \end{center}
\end{titlepage}
\setcounter{footnote}{0}

This letter has been motivated by the recent report on the atmospheric
neutrino by the Superkamiokande collaboration \cite{s-kamioka}, which
shows a clear deficit of muon neutrino \cite{atmos},
\begin{eqnarray}
  \frac{(N_{\nu_\mu}/N_{\nu_e})_{\rm
      data}}{(N_{\nu_\mu}/N_{\nu_e})_{\rm MC}} \,= \left\{
    \begin{array}{ll} 
      0.635 \pm 0.035 \pm 0.053 &\; \mbox{(Sub-GeV)} \\[1.5mm]
      0.604 \pm 0.065 \pm 0.065 &\; \mbox{(Multi-GeV)}
    \end{array} \right..
\end{eqnarray}
The zenith angle distributions of the data are found to be reproduced
quite well by requiring almost maximal mixing between $\nu_\mu$
and $\nu_\tau$ and the following squared mass difference \cite{atmos}:
\begin{equation}
  \sin^2 2 \theta_{\mu\tau} = 0.8 \sim1\,,\qquad \Delta m^2_{23}
  \,=\, (0.4 \sim 5) \times10^{-3} \;{\rm eV}^2.
  \label{atm}
\end{equation}

In addition to these, there is another information on neutrinos if
we take the result of solar neutrino deficit seriously; then, the
following MSW solutions \cite{MSW} are known to resolve the
problem \cite{solar},\footnote{There is another solution with large
  mixing angle which is less preferable in view of the recent
  Superkamiokande reports on the day-night effect and the electron
  recoil energy spectrum \cite{solar}.}
\begin{eqnarray}
  \sin^2 2 \theta_{e\mu} = 0.003 \sim0.012\,,\qquad \Delta m^2_{12}
  \,=\, (0.4 \sim1.1) \times10^{-5} \;{\rm eV}^2. 
   \label{solar}
\end{eqnarray}

The minimum gauge group of the GUT including right-handed neutrino
would be $SO(10)$, in which quarks and leptons appear on the same
footing and, in particular, the neutrino is an $SU(4)_{\rm
  Pati-Salam}$ partner of the up-type quark.  In such a framework, the
Dirac-type masses for the neutrinos and charged leptons should be
parallel with those for the up and down-type quarks,
respectively. Then the additional freedom left in the neutrino mass
matrix is only the right-handed Majorana mass matrix which comes from
the Yukawa couplings of (the standard gauge group) singlet
Higgs,\footnote{If a triplet Higgs exists, its VEV may be induced in a
  class of GUT models and can yield the left-handed Majorana mass term
  directly. This possibility has been examined in many
  literatures\cite{triplet}. We assume here the models in which such
  direct left-handed Majorana mass terms are negligible\cite{CM}.} and
all those together determine the left-handed Majorana masses by
diagonalization procedure \cite{Majorana}.

Now our question is whether those neutrino masses and mixing angles
(\ref{atm}) and (\ref{solar}) are compatible with the ordinary lepton
and quark mass matrices in the GUT framework.

Significant features of the observed CKM matrix and mass eigenvalues
in the quark sector are the quite small mixing and large hierarchical
mass structure. One might reproduce such extremely small mixing by a
strong cancellation between the up and down sectors making use of,
e.g., the so-called ``democratic'' mass matrices \cite{demo}\@. However
we want to consider a more natural scenario in which small parameters
appear without any fine-tuning, and adopt the most economical texture
for the down-type quark and charged lepton sector (Georgi-Jarlskog
type \cite{GJ}):
\begin{eqnarray}
  M_d = m_b \pmatrix{
    0           &   \lambda^3   &      0       \cr
    \lambda^3   &   \lambda^2   &   \lambda^2  \cr
    0           &   \lambda^2   &      1           }, \qquad
  M_l = m_\tau\pmatrix{
    0          &     \lambda^3  &     0        \cr
    \lambda^3  &  -3 \lambda^2  &   \lambda^2  \cr
    0          &     \lambda^2  &     1            },
  \label{d-l}
\end{eqnarray}
with $\lambda= \sin \theta_{\rm C} \simeq 0.22$. This beautifully
reproduces all the masses of the down-type quarks and charged leptons
as well as the CKM mixing except for the 1-3 mixing angle which is
supposed to come from the up quark sector.

As for the lepton sector, the mixing matrix is written as
\begin{equation}
  V_{\rm KM} \,=\, U_l^\dagger U_\nu
  \label{KM}
\end{equation}
where
\begin{equation}
  U_l^\dagger M_l V_l \,=\, {\rm diag}(m_e, m_\mu, m_\tau)\,,\qquad
  U_\nu^\dagger M_\nu U_\nu\,=\, {\rm diag}(m_{\nu_1}, m_{\nu_2 },
  m_{\nu_3} )\,.
\end{equation}
The left-handed neutrino mass matrix $M_\nu$ is given by
\begin{equation}
  M_\nu= M_{\nu{\rm D}}^{\rm \,T}\, M_{R}^{-1} M_{\nu{\rm D}},
  \label{Mnu}
\end{equation}
in the leading order in the inverse power $M_{R}^{-1}$ of the
right-handed neutrino mass matrix $M_{R}$, and $M_{\nu{\rm D}}$ is the
neutrino Dirac mass matrix which is directly related to the up quark
mass matrix which we shall study shortly.

First of all we show how the small mixing angle solution,
$\sin^2 2\theta_{e\mu} = 0.003 \sim0.012$ (corresponding to 
$\sin \theta_{e\mu} = 0.027 \sim0.055$), can be consistent with the
Cabibbo angle.  At first sight one might naively think from the form
(\ref{d-l}) of $M_l$ that $\sin \theta_{e\mu} \simeq
\lambda^3/3\lambda^2 = \lambda/3 \sim 0.073$, which is still larger
and outside the allowed region of the MSW small angle
solution \cite{FY}\@. But note that $V_{\rm KM}$ consists of two parts,
$U_l$ and $U_\nu$, and that this calculation is based on the
presumption that $U_\nu\simeq 1$ just like the up quark sector
$U_u$. Now, the Superkamiokande report requires the maximal
$\nu_\mu\hbox{--}\nu_\tau$ mixing, which in our scenario should come
from the $U_\nu$ side. Then the expression for the KM matrix should
read
\begin{eqnarray}
  V_{\rm KM} \,=\, U_l^\dagger U_\nu\,\sim\,
  \pmatrix{
    1           &  \lambda/3   &     0       \cr
    -\lambda/3  &     1        &  -\lambda^2 \cr
    0           &  \lambda^2   &     1       \cr}
  \pmatrix{
    1     &       0       &      0      \cr
    0     &   1/\sqrt{2}  &  1/\sqrt{2} \cr
    0     &  -1/\sqrt{2}  &  1/\sqrt{2} \cr},
\end{eqnarray}
which predicts $\,\sin \theta_{e\mu} \simeq \lambda/(3\sqrt{2}) \simeq
0.052$, namely, $\sin^2 2\theta_{e\mu} \simeq 0.011$. This is well
within the desired range of the MSW small angle solution. It is really
remarkable that the only input of the maximal
$\nu_\mu\hbox{--}\nu_\tau$ mixing leads to a natural resolution for
the discrepancy between the Cabibbo and MSW angles.

Next we show how the above $\nu_\mu\hbox{--}\nu_\tau$ maximal mixing
can be compatible with the GUT relations which connect the up-type
quark mass matrix to the neutrino Dirac mass matrix. There are two
types of economical texture to reproduce the low-energy up-type quark
mass matrix when the down-type one is given by (\ref{d-l}).
\begin{itemize}
\item type A
  \begin{equation}
    M_u \,=\, m_t \pmatrix{
      x^2   &   0   &   0   \cr
      0     &   x   &   0   \cr
      0     &   0   &   1        },
  \end{equation}
\item type B  \cite{up}
  \begin{equation}
    M_u \,=\, m_t \pmatrix{
      0    &   0    &   x   \cr
      0    &   x    &   0   \cr
      x    &   0    &   1   \cr},
  \end{equation}
\end{itemize}
where $x$ is a small parameter of order $\sim({1/100}-{1/300})\sim
m_c/m_t$. The neutrino Dirac mass matrix $M_{\nu{\rm D}}$ is given in
quite a parallel way with the above texture by replacing $m_t$ with
a scale $m \sim m_t/3$ \ in which the factor 3 represents the effect
of renormalization group equation (RGE).

Using eq.\ (\ref{Mnu}), we can get the left-handed Majorana mass
matrix once the right-handed Majorana mass matrix $M_R$ is given. Our
task is to seek for the texture of $M_R$ which can reproduce the
present experimental values of (\ref{atm}) and (\ref{solar}). In view
of the mass difference data, $\Delta m^2_{12}$ and $\Delta m^2_{23}$,
the most natural possibility for the left-handed neutrino mass pattern
is the following hierarchical one:
\begin{eqnarray}
  m_{\nu_1} \ll m_{\nu_2} \ll m_{\nu_3}\,.
  \label{hierarchy}
\end{eqnarray}
Indeed, in this case, the squared mass difference data just tells us
$\Delta m_{23}^2 \simeq m_{\nu_3}^2$ and
$\Delta m_{12}^2 \simeq m_{\nu_2}^2$
so that $m_{\nu_3} \sim (2-7)\times 10^{-2}\,$eV and $m_{\nu_2} \sim
(2-3) \times 10^{-3}\,$eV\@. There is no fine-tuning required and so
this is a natural possibility.

We here adopt the case of type B for $M_{\nu{\rm D}}$, since it
realizes more natural pattern for the right-handed neutrino mass
matrix than the type A case as we shall see. The characteristic
feature of this type B texture is that the generation labels are
exchanged between the first and the third and we shall call it ``label 
exchanging texture''. We can obtain a large $\nu_\mu\hbox{--}\nu_\tau$
mixing angle together with the hierarchical mass pattern
(\ref{hierarchy}) by taking the following texture for the right-handed 
Majorana mass matrix with $O(1)$ coefficients $\alpha,\beta,\gamma$.
\begin{eqnarray}
  M_R \,=\, \pmatrix{
    \alpha M  &  \beta M   &   0   \cr
    \beta M   &  \gamma M  &   0   \cr
    0         &     0      &   M'     \cr}.
\end{eqnarray}
The resultant left-handed Majorana mass matrix becomes
\begin{eqnarray}
  M_\nu &=& M_{\nu{\rm D}}^{\rm \,T}\, M_R^{-1} M_{\nu{\rm D}}
  \nonumber \\[3mm]
  &=& m^2 \pmatrix{
    0   &   0   &   x   \cr
    0   &   x   &   0   \cr
    x   &   0   &   1      }
  \pmatrix{
    a M^{-1}  &  b M^{-1}  &    0      \cr
    b M^{-1}  &  c M^{-1}  &    0      \cr
    0         &     0      &  M'^{-1}     }
  \pmatrix{
    0   &   0   &   x   \cr
    0   &   x   &   0   \cr
    x   &   0   &   1      }  \nonumber  \\[3mm]
  &=& \frac{m^2 x^2}{M} \pmatrix{
    \epsilon x^2  &   0   &  \epsilon x   \cr
    0             &   c   &      b        \cr
    \epsilon x    &   b   &  a + \epsilon \cr }
  \label{leftmass} \\[3mm]
  && {\rm with} \;\;\pmatrix{a & b \cr b & c} \equiv
  \pmatrix{\alpha & \beta \cr \beta & \gamma}^{-1}, \qquad
  \epsilon \equiv \frac{M}{M'x^2}\,.
\end{eqnarray}
In order to get a large $\nu_\mu\hbox{--}\nu_\tau$ mixing angle, say,
$\left| \tan \theta - 1 \right| \lsim 0.2$, the condition $\left|
  (a+\epsilon-c)/b \right| \lsim 0.4$ should be satisfied. So, in
particular, 
\begin{equation}
  \epsilon \ \lsim \ O(1)
\end{equation}
should hold since $a,\,b,\,c$ are of order one. For the mass hierarchy 
$m_{\nu_2}/m_{\nu_3}\ll 1$ to be realized, the following determinant
should be small:
\begin{equation}
  \frac{1}{(a+\epsilon+c)^2} \det \pmatrix{c & b \cr b &
    a+\epsilon \cr} \ \sim\ \frac{m_{\nu_2}}{m_{\nu_3}}\,.
\end{equation}
To see the situation more concretely, we can take, for example,
\begin{eqnarray}
  \epsilon = x, \quad a = c = 1.4\,,\quad b = 1.2,
\end{eqnarray}
which corresponds to $\alpha = \gamma \simeq 2.7$ and $\beta \simeq
-2.3$. Then, using $m_{\nu_3} \sim 3.6 \times 10^{-2}\,{\rm eV}$ (the
best fit value for the atmospheric neutrino anomaly \cite{atmos}) as an
input together with $m \sim m_t/3 \sim 60\,$GeV, we obtain, for a
range of $x$ values, $x \sim1/100 - 1/300$, the right-handed neutrino
masses
\begin{eqnarray}
  \alpha M \ \sim\  \left|\beta\right|M &\sim& (7.0 - 0.7) \times
  10^{10} \;{\rm GeV} \\
  M'\ = \ M/x^3 &\sim& (2.6 - 7.8) \times 10^{16} \;{\rm GeV},
\end{eqnarray}
and the left-handed neutrino mass eigenvalues and the mixing matrix
\begin{eqnarray}
  m_{\nu_1} &\sim& (1.3\times 10^{-8} - 5.0\times 10^{-10}) \;{\rm
    eV} \,, \\
  m_{\nu_2} &\sim& 2.8 \times10^{-3} \;{\rm eV} \,, \\
  m_{\nu_3} &\sim& 3.6 \times10^{-2} \;{\rm eV} \ {\rm (input)}\,,
\end{eqnarray}
\begin{eqnarray}
  V_{\rm KM} &\simeq& \pmatrix{
    0.997   &   0.049  &  0.049  \cr
    -0.069  &   0.736  &  0.675  \cr
    -0.0001 &  -0.677  &  0.737     } .
  \label{mixing}
\end{eqnarray}
The mixing matrix elements are stable up to $\lsim 1/1000$ in this
range of $x$. These values correspond to 
\begin{eqnarray}
  \sin^ 2 2 \theta_{e\mu} &\!\!\sim\!\!& 0.0094 \,,\quad \Delta
  m^2_{12} \,\sim\, 0.8 \times10^{-5} \;{\rm eV}^2  \,, \\[1.5mm]
  \sin^2 2 \theta_{\mu\tau} &\!\!\sim\!\!& 1 \,,\hspace*{13.5mm}
  \Delta m^2_{23} \,\sim\, 1.3 \times10^{-3} \;{\rm eV}^2 \,,
\end{eqnarray}
and are surely in agreement with the recent experimental data
(\ref{atm}) and (\ref{solar}).

It is interesting to note that in this label-exchanging texture case
the right-handed Majorana mass of the third generation becomes the
order of GUT scale: $M' \simeq 10^{16} \,{\rm GeV} \simeq M_{\rm
  GUT}$. This feature is quite favorable for the GUT scenario to be
consistent with the observed bottom-tau mass ratio; in the usual MSSM,
the right-handed neutrino with intermediate scale mass suppresses the
effect of top Yukawa contribution to RGE and enhances the bottom-tau
mass ratio substantially, and it is required that the right-handed
Majorana mass of the third generation must be heavier than at least
$10^{13}\,$GeV \cite{b-tau}. If this label-exchanging type seesaw
mechanism is realized, the right-handed neutrino of the third
generation can be far heavier than the other two which are of the
order of ordinary intermediate scale
$\sim10^{10}\>$GeV\@. Interestingly enough, this right-handed neutrino
of the third generation essentially determines the light neutrino mass
of the first generation on the one hand, and contributes to the
improvement for the problem of the fermion masses of the third
generation, i.e., the bottom-tau mass ratio problem on the other
hand. This can occur only in the type B texture for the up-type quark
mass matrix.

We should add a comment on the fact that there is
{\it no physical significance} in reproducing the desired form for the
neutrino mass matrix $M_\nu$ itself. This is because, for any given
Dirac mass matrix $M_{\nu{\rm D}}$, we can reproduce any desired form
for the neutrino mass matrix $M_\nu$ by adjusting the right-handed
neutrino mass matrix $M_R$; namely, we can take $M_R = M_{\nu{\rm D}}
M_\nu^{-1} M_{\nu{\rm D}}^{\rm T}$. But the physics resides in the
form of $M_R$. If the resultant form of $M_R$ suggests a {\it natural} 
physics for the right-handed neutrino sector which is unknown yet,
then the form of the original Dirac mass term $M_{\nu{\rm D}}$ will be
meaningful. For instance, consider the type A Dirac mass matrix 
for $M_{\nu{\rm D}}$. Then (essentially) the same form of the
left-handed neutrino mass matrix $M_\nu$ as eq.\ (\ref{leftmass})
can be realized by taking
\begin{eqnarray}
  M_R \,=\, \pmatrix{
    M'x^2        & -\beta M   &  -\alpha M/x    \cr
    -\beta M     & \gamma M   &  \beta M/x      \cr
    -\alpha M/x  & \beta M/x  &  \alpha M/x^2   \cr}.
\end{eqnarray}
This form is, however, not very illuminating, and moreover contains
three different scales for the right-handed neutrino masses:
$M\sim10^{10}\,$GeV,\ $M'x^2 \sim M/x \sim 10^{12}\>$GeV, \  $M/x^2
\sim 10^{14}\>$GeV\@. So, in this sense, the type A Dirac mass matrix
is not preferable.

A few more comments are in order: 

The phenomenological nature of the present work should be noted: what
we have done in this short note was just to have proposed a simple
neutrino texture with large mixing as an extract of the neutrino
data. But, as we have emphasized, it has natural features very
suggestive for the existence of a simple underlying unified theory. It
would, therefore, be an interesting and important next step to
construct concrete models realizing this neutrino texture in a natural
way.

The recent results of the CHOOZ long-baseline neutrino oscillation
experiment \cite{CHOOZ} imply that the $\nu_e\hbox{--}\nu_\tau$ mixing
is small if one assumes the oscillations between three neutrino
flavors and the hierarchical mass pattern \cite{Ue3}\@. This is in good
agreement with our texture and resultant mixing angles
(\ref{mixing}). However, our texture is incompatible with the recent
LSND observations \cite{LSND}\@. It seems difficult to explain all the
above neutrino oscillation experiments within the models with three
species of neutrinos. If we take the cosmological indication of the
neutrino with mass of a few eV as a candidate for the hot dark matter,
we can also introduce the additional sterile neutrino other than the
three neutrinos, since our texture predicts all the three neutrino
masses are smaller than 0.1 eV\@. For models accommodating the hot
dark matter results, see the papers \cite{DM}. 

In conclusion, we have shown that there is a good candidate for the
texture of right-handed Majorana mass matrix to generate the large
$\nu_\mu\hbox{--}\nu_\tau$ mixing with hierarchical masses. This
texture reproduces the recent Superkamiokande measurements quite
naturally as well as the MSW small angle solution for the solar
neutrino problem without any fine-tuning. A remarkable fact is that we
can adopt the charged lepton and neutrino mass matrices which is quite
consistent with the GUT relations to the quark sector, and especially
we found that the small angle solution of MSW can be explained by the
Cabibbo angle only, if the mixing angle for $\nu_\mu\hbox{--}\nu_\tau$
is very large.

\section*{Acknowledgements}
We started this work by a stimulating discussion with T.\ Yanagida
while the Summer Institute '97 held at Yukawa Institute in August
1997. We are also indebted to him for his encouragement by many
helpful discussions and his kindness to give us valuable new
information on Superkamiokande. Discussions with N.\ Maekawa and J.\
Sato are also appreciated. M.\ B. and T.\ K. are supported in part by
the Grant-in Aid for Scientific Research No. 09640375 and 08640367,
respectively.

\end{document}